\pdfoutput=1

\documentclass[%
english,
 reprint,
 superscriptaddress,
showpacs,preprintnumbers,
nofootinbib,
 amsmath,amssymb,
 aps,
prd,
]{revtex4}

\usepackage{graphicx}
\usepackage{dcolumn}
\usepackage{bm}

\usepackage{subfigure}
\usepackage{multirow}
\usepackage{epsf}

\newcommand{\eqn}[1]{eq.~(\ref{#1})}

\newcommand{\Eqn}[1]{Eq.~(\ref{#1})}

\newcommand{\gsim}
{\mbox{${~\raise.25em\hbox{$>$}\kern-.70em
\lower.25em\hbox{$\sim$}~}$}}
\newcommand{\lsim}
{\mbox{${~\raise.25em\hbox{$<$}\kern-.70em
\lower.25em\hbox{$\sim$}~}$}}
\newcommand{\beq}{\begin{equation}}
\newcommand{\beqa}{\begin{eqnarray}}
\newcommand{\eeq}{\end{equation}}
\newcommand{\eeqa}{\end{eqnarray}}

\begin{document}

\title{Quark masses, mixings, and CP violation
 from spontaneous breaking of flavor $SU(3)^{3}$ }

\author{Chee Sheng Fong}
\email{chee.sheng.fong@lnf.infn.it} 
\affiliation{INFN, Laboratori Nazionali di Frascati 
   CP 13, I00044 Frascati, Italy}  

\author{Enrico Nardi} 
\email{enrico.nardi@lnf.infn.it} 
\affiliation{INFN, Laboratori Nazionali di Frascati  
   CP 13, I00044 Frascati, Italy}

\begin{abstract}
  A ${\cal G}_{\cal F}=SU(3)_{Q}\times SU(3)_{u}\times SU(3)_{d}$
  invariant scalar potential breaking spontaneously the quark flavor
  symmetry can explain the standard model flavor puzzle. 
  The approximate alignment in flavor space of the vacuum expectation values of the
  up and down ``Yukawa fields'' is explained as a dynamical effect, and
  the observed quark mixing angles, the weak CP violating phase, and
  hierarchical quark masses can be reproduced without introducing
  hierarchical parameters.
\end{abstract}



 \maketitle


\section{Introduction}

Although the standard model (SM) is an extremely successful theory in
describing the fundamental building blocks of our Universe, it
contains some unpleasant features, among which a puzzling flavor
structure, which is characterized by large hierarchies between Yukawa
couplings and by a quark mixing matrix that, without any apparent
reason, is approximately proportional to the identity.  In the absence
of Yukawa interactions, the SM has a large global symmetry
$G=U(3)_{Q}\times U(3)_{u}\times U(3)_{d}\times U(3)_{\ell}\times
U(3)_{e}\times U(1)_H$~\cite{Chivukula:1987py} where the subscripts
refer respectively to the quark $SU(2)$ doublets $Q$ and up and down
singlets $u$, $d$, lepton doublets $\ell$ and singlets $e$, and to the
Higgs field $H$.  When Yukawa interactions are turned on, the
surviving symmetry reduces to $U(1)_Y \times U(1)_{B}\times \Pi_\alpha
U(1)_{L_\alpha}$ corresponding to hypercharge, baryon, and lepton
flavor numbers, which are linear combinations of $U(1)_H$ and of the
Abelian factors contained in the five $U(3)=SU(3)\otimes U(1)$.  Of
course, the lepton flavor symmetries are broken by the mixing between
massive neutrinos. However, describing this breaking requires some
hypothesis about the specific extension of the SM model responsible
for neutrino masses, which is not unique.  Therefore, in this paper we
will concentrate on the quark sector, for which the flavor symmetry
breaking pattern is fully contained within the SM, and experimentally
known with good accuracy.  Following a general approach~
\cite{Anselm:1996jm,%
  Berezhiani:2001mh,Koide:2008qm,Koide:2008tr,Koide:2012fw,Feldmann:2009dc,%
  Albrecht:2010xh,Grinstein:2010ve,Alonso:2011yg,Nardi:2011st,Alonso:2013nca},
we assume that at some large energy scale the quark flavor symmetry $
{\cal G}_{\cal F}=SU(3)_{Q}\times SU(3)_{u}\times SU(3)_{d}\subset G$
is an exact (global or local) symmetry of nature.\footnote{We will not
  be interested here in the Abelian factors contained in $G$. Some
  possible roles of these factors are analyzed for example
  in~\cite{Nardi:2011st} for the quarks and in~\cite{Alonso:2011jd}
  for the leptons.}  We are interested in the spontaneous breaking of
${\cal G}_{\cal F}$ which occurs when the scalar ``Yukawa fields''
coupled to quark bilinears via dimension-five operators, and
transforming nontrivially under the symmetry, acquire vacuum
expectation values (vevs) which give rise to the Yukawa couplings we observe.

The effective dimension-five Yukawa operators read
\begin{eqnarray}
\label{eq:Ly}
-{\cal L}_{Y} & = & 
\sum_{q=u,d} \left[\,\frac{\tilde\kappa_q}{\Lambda}\, \overline{Q}\,Y_{q}
\,q\, H_q+{\rm h.c.}\,\right],
\end{eqnarray}
where $H_d=H$ is the Higgs field and $H_u=i\sigma_2 H$,
$\tilde\kappa_{u,d}$ are dimensionless (complex) couplings, $Y_{u,d}$
are the up- and down-type Yukawa fields, and $\Lambda$ is the high
scale where the effective operators arise.\footnote{We neglect
  effective operators of dimension higher than five.  This is
  justified for example if $\Lambda\gsim10^{9}$ GeV  
  (as is needed in case the flavor symmetry is global, to suppress 
  sufficiently the flavor changing neutral current (FCNC) couplings  
  of the Nambu-Goldstone bosons with quarks) 
which corresponds to a top-quark coupling  $y_{t}\sim\left\langle
    Y_{u,33}\right\rangle /\Lambda\lsim0.6$ \cite{Xing:2007fb}.}  At
lower energy the flavor symmetry gets broken, and the SM Yukawa
couplings ${\cal Y}_{q}=\tilde\kappa_q\left\langle Y_{q}\right\rangle
/\Lambda$ are eventually generated.

Of course, the theoretical challenge is to find a ${\cal G}_{\cal F}
$-invariant scalar potential $V(Y_q,Z)$ (where $Z$ denotes generically
additional scalars coupled to $Y_q$ in a symmetry invariant way)
which can spontaneously break ${\cal G}_{\cal F} $ yielding a set of
vevs $\langle Y_{q}\rangle$ with the observed structure of the SM
Yukawa couplings.

%
%

In ref.~\cite{Nardi:2011st} it was found that the most general
renormalizable ${\cal G}_{\cal F} $-invariant potential with only
$Y_u$ and $Y_d$ admits the tree-level vacuum configuration $\langle
Y_{q}\rangle \sim {\rm diag}(0,0,v_{q})$. This appeared as a promising
starting point to account for the hierarchies $m_t\gg m_{c,u}$ and
$m_b \gg m_{s,d}$ . However, in ref.~\cite{Espinosa:2012uu} it was
proven that the vanishing entries in $\langle Y_q\rangle $ cannot be
lifted to nonvanishing values by any type of perturbative effects
(loop corrections or higher dimensional operators involving $Y_{u,d}$
only).\footnote{Stated in another way, this implies that no type of
  perturbative effect can further break the little groups left
  unbroken by the minima of the tree-level potential.  We will then
  assume that nonrenormalizable operators remain subdominant with
  respect to the renormalizable ones, in which case neglecting them 
  leaves our results qualitatively unchanged.}  Nevertheless, in
ref.~\cite{Espinosa:2012uu} it was also shown that by including
additional scalar multiplets transforming in fundamental
representations of the $SU(3)$ factors of ${\cal G}_{\cal F}$, a
scalar potential which admits a hierarchical ground state $\langle
Y_q\rangle \sim v_q\, {\rm diag}\left(\epsilon',\epsilon,1\right)$
with $\epsilon'\ll\epsilon\ll1$ can be constructed.

In this work, we show that realistic quark masses and 
Cabibbo-Kobayashi-Maskawa (CKM) mixing \cite{Cabibbo:1963yz,Kobayashi:1973fv} 
as well as the weak CP violating phase can indeed be obtained from the
spontaneous breaking of ${\cal G}_{\cal F}$.  In section 2 we
introduce a general classification of the different field monomials
that can appear in $V(Y_q,Z)$.  In section 3 we rederive the
hierarchical solution $\langle Y_q\rangle \sim v_q\, {\rm
  diag}\left(\epsilon',\epsilon,1\right)$ obtained in
ref. \cite{Espinosa:2012uu}.  In section 4 we discuss which is the
minimal field content needed to obtain nontrivial quark mixings.
In section 5 we write down the most general ${\cal G}_{\cal
  F}$-invariant potential for this minimal field content, and we show
that it automatically implies that one weak CP violating
phase is generated at the potential minimum.  
In section 6 we discuss briefly a numerical example which produces realistic
quark masses, mixing angles and the weak CP violating phase.  The counting
of physical complex parameters in $V(Y_q,Z)$ is carried out in the
Appendix.

\section{Generalities and notations}

Constructing a ${\cal G}_{\cal F}$-invariant potential which can yield
at its minimum a symmetry breaking pattern with Yukawa vevs $\langle
Y_q \rangle$ in agreement with observations requires, besides the
Yukawa fields $Y_{u,d}$, the inclusion of additional scalars that we
generically denote with $Z$. The resulting potential $V(Y_q,Z)$
contains various operators describing interactions and
self-interactions between the different fields which, by themselves,
tend to break ${\cal G}_{\cal F}$ in some specific way. It is
then useful to introduce a classification of these operators based  
on  their dynamical properties with respect to minimization. 

Let us start, as a first example, with the $SU(3)_Q\times SU(3)_q$
invariant potential for a single Yukawa field $Y_q \sim
\left(\mathbf{3},\mathbf{\bar 3}\right)$.  We have three invariants:
\begin{eqnarray}
T_{q} & = & {\rm Tr}\left(Y_{q}Y_{q}^{\dagger}\right),
\nonumber \\
A_{q} & = & \frac{1}{2}\left[T_{q}^{2}-
{\rm Tr}\left(Y_{q}Y_{q}^{\dagger}Y_{q}Y_{q}^{\dagger}\right)\right],
\nonumber \\
{\cal D}_{q} & = & \det Y_{q}, 
\end{eqnarray}
and the  scalar potential $V(Y_q)$  is~\cite{Nardi:2011st} 
\begin{eqnarray}
\label{eq:VYq}
V(Y_q)&=& V_{\cal I} + V_{\cal AR} + V_{\cal A}\,, \\
\label{eq:Vl}
V_{\cal I}&=&\lambda\left[T_q-v^2_q\right]^2,\\
\label{eq:Vf}
V_{\cal AR}&=&\lambda_A A_q\,, \\
\label{eq:Vfp}
V_{\cal A} &=& \tilde \mu_q\, {\cal D}_q + {\rm h.c.} 
= 2\mu_q\,D_q\, \cos \delta_{q} \,,
\end{eqnarray}
where in the last line we have introduced $\mu_q = |\tilde \mu_q|$,
$D_q=|{\cal D}_q|$, and $\delta_{q}= {\rm Arg}\, {\cal D}_q$ (the
phase of $\mu_q$ can always be reabsorbed by redefining $\delta_q$,
see below). \footnote{ As long as
  $\langle H^\dagger H \rangle/\Lambda^2 \ll 1$ the coupling with the
  Higgs, $ H^\dagger H T$, can be omitted from \eqn{eq:VYq}. Regarding
  the effects of such coupling on the Higgs potential, electroweak
  symmetry breaking at the correct scale would require a certain
  degree of fine-tuning in the term $H^\dagger H\left(\langle
    T\rangle-\mu_H^2\right)$.}

As discussed in~\cite{Nardi:2011st} this potential allows for two
patterns of breaking the $SU(3)_Q\times SU(3)_q$ symmetry: $\langle
Y_q\rangle^s \sim v_q {\rm diag} (1,1,1)$ which yields the maximal
little group $SU(3)_{Q+q}$, and $\langle Y_q\rangle^h \sim v_q {\rm
  diag}(0,0,1)$ which yields the maximal little group $SU(2)_{Q}\times
SU(2)_{q}\times U(1)$. We will define as {\it attractive} ($\cal A$)
those terms in the potential that tend to break the symmetry to the
largest maximal little group (in this case $SU(3)_{Q+q}$ with eight
generators), and {\it repulsive} ($\cal R$) those terms that tend to
break the symmetry to the smallest maximal little group (in this case
$SU(2)_Q\times SU(2)_{q}\times U(1)$ with seven generators), and as
{\it flavor irrelevant} ($\cal I$) those operators that are blind to
particular configurations of the flavor symmetry breaking minimum.

$T_q$ in $V_{\cal I}$ \eqn{eq:Vl} is an example of a flavor irrelevant
operator.  This is because $T_q$ is invariant under the accidental
symmetry $SO(18)$ which is much larger than the flavor symmetry, and
that gets broken to $SO(17)$ by the vev $\langle T_q\rangle \neq
0$. This operator is then flavor irrelevant because the value of its
vev does not depend on any particular flavor configuration, given
that $SO(17)$ transformations can rotate e.g. $\langle Y_q\rangle^{h}$ into
$ \langle Y_q\rangle^{s}$.  Then the role of $V_{\cal I}$ \eqn{eq:Vl} is just
that of determining the ``length'' of the vev of $Y_q$ (defined as
$\sqrt{T_q}$), while it does not contribute to the determination of
any specific  flavor direction.

$V_{\cal AR}$ in \eqn{eq:Vf} contains the operator $A_q$ which 
corresponds to a Hermitian monomial. $A_q$ contributes to determine the flavor
structure, and it can be both attractive or repulsive: it is
easily seen that for $\lambda_A<0$ the action of $A_q$ is attractive,
since the minimum of the potential is lowered for the largest possible
value of $\langle A_q\rangle$, which is obtained for $\langle
Y_q\rangle = \langle Y_q\rangle^s$. If $\lambda_A>0$, then the action
of $A_q$ is repulsive, since its minimum value $\langle A_q\rangle=0$
is obtained for $\langle Y_q\rangle = \langle Y_q\rangle^h$.

Operators which correspond to non-Hermitian monomials are included in
$V_{\cal A}$. Non-Hermitian monomials are always attractive, as is the
case for ${\cal D}_q$ in~\eqn{eq:Vfp}. For example, when $\langle
{\cal D}_q\rangle$ is nonvanishing, minimization drives its phase
$\delta_q\to \pi$ ( $\cos \delta_q\to -1$).  Then the potential
minimum gets lowered for the largest possible value of $D_q$, which is
obtained for $\langle Y_q\rangle = \langle Y_q\rangle^s$.  Besides
being relevant for determining the flavor structure, in those cases
in which $V_{\cal A}$ contains physical complex phases (see
section~\ref{sec:CP}), it will also have the important role of
determining the value of the SM weak CP violating phase.

For interactions between different fields the jargon {\it attractive},
{\it repulsive}, {\it flavor irrelevant}, acquires a more intuitive
meaning.  Let us introduce for example two additional fields $Z_Q$ and
$Z_q$ transforming under $SU(3)_{Q}\times SU(3)_{q}$ respectively as
$\left(\mathbf{3},\mathbf{1}\right)$ and
$\left(\mathbf{1},\mathbf{3}\right)$.  Also in this case we have terms
that are invariant under extended accidental symmetries, and that are
flavor irrelevant:\footnote{Here and in the rest the paper, we use
  the modulus square notation $|A|^2= A^{\dagger} A$ (where $A$ is
  generically a column vector that can also result from the product of
  vectors and square matrices) to put in evidence the Hermiticity of
  the corresponding monomials.}
\begin{eqnarray}
|Z_Q|^2, \quad |Z_q|^2 \quad\quad\  &\qquad\qquad&  SO(6), \\
|Z_Q|^2\cdot |Z_q|^2\qquad\quad &\qquad\qquad&  SO(6)\times SO(6), \\
T_q |Z_Q|^2,\quad  T_q |Z_q|^2\  &\qquad\qquad& SO(18)\times SO(6).
\end{eqnarray}
We  assign this type of terms to $V_{\cal I}$. 
Hermitian monomials like 
\begin{eqnarray}
\alpha_{q}\,  |Y_{q}Z_{q}|^2, \qquad   
\alpha^Q_q \, |Y_{q}^{\dagger} Z_{Q}|^2    
\end{eqnarray}
can be attractive or repulsive depending if their real couplings are
negative or positive, and are assigned to $V_{\cal AR}$.  Assuming for
example that $Y_q$ acquires the vev $\langle Y_q\rangle^h \sim v_q
{\rm diag}(0,0,1)$, we see that $\alpha^Q_q<0$ favors the aligned
configuration $\langle Z_{Q}^T \rangle \sim v_{Z_Q}\left(0,0,1\right)$ because it
maximizes the vev $\langle\, |Y_{q}^{\dagger}Z_{Q}|^2\, \rangle $,
while $\alpha^Q_q>0$ would favor the orthogonal configuration 
$\langle Z_{Q}^T \rangle  \sim v_{Z_Q}\left(c,s,0\right)$ (with $c^2+s^2=1$) which yields
$\langle\, |Y_{q}^{\dagger}Z_{Q}|^2\, \rangle =0 $. 
Operators corresponding to non-Hermitian monomials are 
included in $V_{\cal A}$. An example is 
\begin{eqnarray}
\tilde\nu_q\, Z_Q^\dagger Y_{q}Z_{q} + {\rm h.c.} = 2\, \nu_q\, \left| 
Z_Q^\dagger Y_{q}Z_{q}\right| \, \cos{\phi}, 
\end{eqnarray}
where $\nu_q=|\tilde \nu_q|$ while $\phi$ denotes the overall phase of
the term.  These operators are always attractive, since at the minimum
$\phi \to \pi$ and thus they always give a negative contribution to
the potential.

It should be clear by now that attractive terms (non-Hermitian
monomials and Hermitian monomials with negative couplings) tend to
align in flavor space the vevs of different multiplets, while
repulsive terms (Hermitian monomials with positive couplings) favor
``orthogonal'' or more generically ``maximally misaligned'' vevs
configurations.\footnote{According to the standard usage, ``orthogonal''
  describes the situation in which the product of two vevs vanishes
  exactly. With `maximally misaligned' we will instead refer to the
  situation in which, at fixed lengths, the product of two or more
  vevs is made as small as possible in absolute value.}

\section{Hierarchical Yukawa couplings}
\label{sec:hierarchy}

We start by describing, following ref.~\cite{Espinosa:2012uu} , which
type of scalar potential is needed to generate a vev $\langle
Y_q\rangle \sim v_q\, {\rm diag}\left(\epsilon',\epsilon,1\right)$
with hierarchical entries $\epsilon'\ll\epsilon\ll1$.  Focusing on
just one type of quark ($q=u$ or $d$) the flavor symmetry is
$SU(3)_{Q}\times SU(3)_{q}$ under which $Y_q$ transforms as a
bifundamental representation
$Y_{q}\sim\left(\mathbf{3},\mathbf{\bar{3}}\right)$.  Additional
fields are needed to generate solutions different from $\langle
Y_q\rangle^{h,s}$~\cite{Espinosa:2012uu} and thus we add two scalar
multiplets transforming respectively in the fundamental of each one of
the two $SU(3)$ factor $Z_{Q}\sim\left(\mathbf{3},\mathbf{1}\right)$
and $Z_{q}\sim\left(\mathbf{1},\mathbf{3}\right)$.\footnote{Through
  operators of dimension six or higher, the vevs of $Z_{Q,q}$ can give
  rise to new FCNC effective operators.  They can however be forbidden
  by a suitable choice of the representations of the messenger
  fermions that generate the effective operators \eqn{eq:Ly}.} 
With this field content, the most general renormalizable scalar
potential invariant under $SU(3)_{Q}\times SU(3)_{q}$ is
\begin{equation}
V(Y_q,Z_q,Z_{Q}) =V_{\cal I}+V_{\cal AR}+V_{\cal A},
\end{equation}
where
\begin{eqnarray}
V_{\cal I} & = & \lambda_{q}\left(T_{q}-v_{q}^{2}\right)^{2}+\lambda_{Q}
\left(\left|Z_{Q}\right|^{2}-v_{Z_{Q}}^{2}\right)^{2}+\lambda_{Z_{q}}
\left(\left|Z_{q}\right|^{2}-v_{Z_{q}}^{2}\right)^{2}\nonumber \\
 &  & +\left[g_q\left(T_{q}-v_{q}^{2}\right)+g_{Q}
\left(\left|Z_{Q}\right|^{2}-v_{Z_{Q}}^{2}\right)+g_{Z_q}
\left(\left|Z_{q}\right|^{2}-v_{Z_{q}}^{2}\right)\right]^2,
\label{eq:Vlh} \\ 
V_{\cal AR} & = & \lambda_{A_q}A_{q}+
\alpha_{q}\, \left|Y_{q}Z_{q}\right|^2 + 
\alpha^Q_{q}\, \left|Y_{q}^{\dagger}Z_{Q}\right|^2\,, 
\label{eq:Vfh} \\
 V_{\cal A} &= & \tilde{\mu}_{q}{\cal D}_{q}
+\tilde{\nu}_{q}Z_{Q}^{\dagger}Y_{q}Z_{q}+{\rm h.c..}
\label{eq:Vfph}
\end{eqnarray}
The way $V_{\cal I}$ is written makes it clear that it is flavor
irrelevant, and only determines the `lengths' of $T_q,\,Z_Q$ and $Z_q$
at the minimum.  In the following we adopt the convention of denoting
complex quantities with a tilde: $\tilde x$, while the modulus and the
phase of $\tilde x$ will be denoted respectively with $x$ and
$\phi_x$.  The two phases $\phi_{\mu_q}$ and $\phi_{\nu_q}$ in
\eqn{eq:Vfph} can be removed by redefining the fields:
\begin{eqnarray}
Y_{q} & \to & e^{-i\phi_{\mu_{q}}/3}Y_{q},\nonumber \\
Z_{q} & \to & e^{i(\phi_{\mu_q}/3-\phi_{\nu_q})}Z_{q},
\end{eqnarray}
so that the potential is manifestly CP-invariant.
%
%
%
In order to study the ground state, we take all the fields to be
background classical fields (i.e. spacetime independent). However, to
avoid overcluttering the notations, we keep using the same symbols
$Y=Y^c,\, Z=Z^c$ as for the spacetime dependent fields $Y=Y(x),\,
Z=Z(x)$, since the difference should be clear from the context.  We
can make use of the $SU(3)_{Q}\times SU(3)_{q}$ symmetry to choose,
without loss of generality, a convenient basis by removing $3+3$
moduli and $5+5$ phases.  The generic $3\times 3$ matrix $Y_{q}$ has nine
moduli and nine phases.  By rotating away six moduli and six phases it can be
brought to diagonal form, which we will denote by $\hat{Y}_{q}$. Two
additional transformations generated by $\lambda_3^{Q-q}$ and
$\lambda_8^{Q-q}$ (where $\lambda_{3,8}$ are the diagonal $SU(3)$
Gell-Mann matrices) allow one to remove two phases from two diagonal
entries. However, since $\lambda_3^{Q+q}$ and $\lambda_8^{Q+q}$ both
leave $\hat{Y}_{q}$ invariant, the third phase cannot be removed.  We
can however make use of these two $U(1)$ symmetries to remove one
phase from $Z_{Q}$ and another one from $Z_{q}$.  For example, a valid
choice of basis is $\hat{Y}_{q}={\rm
  diag}\left(y_{q}^1,y_{q}^2,\tilde{y}_{q}^3\right)$,
$Z_{Q}^{T}=\left(\widetilde{z}_{Q}^1,\tilde{z}_{Q}^2,z_{Q}^3\right)$
and $Z_{q}^{T}=\left(z_{q}^1,\tilde{z}_{q}^2,\tilde{z}_{q}^3\right)$
where the tilde denotes complex components while the others are real
and positive.

Let us first study whether CP violation can occur spontaneously, that
is if at the minimum, the vevs of some of the remaining five phases are
forced to acquire a value $\neq 0,\, \pi$.  $V_{\cal A}$ in
\eqn{eq:Vfph} is the sum of four pairs of complex conjugate terms which
depend on different combinations of the five phases. Then their
optimal minimization ({\it i.e.} all $\phi \to \pi$) can all be
satisfied by fixing the value of three phases and of one phase
difference. For example, for the choice of basis given above we obtain
$\phi_{y_{q}^3} = \phi_{z_{Q}^1} = \phi_{z_{q}^2}-\phi_{z_{Q}^2} = \pi$
and $\phi_{z_{q}^3}=0$. Thus, the CP conserving potential
$V(Y_q,Z_q,Z_Q)$ also yields a CP conserving minimum.

To study the possible flavor configurations of the minima, we can now
take the potential of classical background fields to be a function of
real parameters and of real classical fields, after accounting for a
minus sign in front of the four pairs of complex conjugate terms in
\eqn{eq:Vfph} which then read:
\begin{equation}
V_{\cal A}^{\rm min} =  -2\mu_{q}\,\Pi_i\,  y_{q}^i
- 2\nu_{q}\sum_i\, z_{Q}^i z_{q}^i y_{q}^i\,. 
\label{eq:Vfph2}
\end{equation}
As already said, $V_{\cal I}$ in~\eqn{eq:Vlh} fixes the length
$\sqrt{T_{q}}= v_{q}$.  We chose $\lambda_{A_{q}}>0$ in \eqn{eq:Vfh}
so that the first term is repulsive and favors the configuration
$\hat{Y}_{q}^h \simeq v_q\, {\rm diag}\left(0,0,1\right)$
\cite{Nardi:2011st}, which is the starting point for generating a
hierarchical solution.  Choosing $\alpha_{q}>0$ and $\alpha_{q}^Q>0$
implies that the last two terms in $V_{\cal AR}$ are also repulsive,
and tend to generate a maximal misalignment between $Z_{Q}$, $Z_{q}$
and $\hat{Y}_{q}$, which  corresponds to $Z_{Q}^{T}= v_{Z_{Q}}
\left(c_{Q},s_{Q},0\right)$ and $Z_{q}^{T}= v_{Z_{q}}
\left(c_{q},s_{q},0\right)$ with $c^2_{Q}+s^2_{Q}=c^2_{q}+s^2_{q}=1$.
On the other hand, the terms proportional to $\nu_q$ in \eqn{eq:Vfph2}
which are always attractive prefer to align $Z_{Q}$ and $Z_{q}$ with
$\hat{Y}_{q}$ in order to get a nonvanishing (and possibly large)
negative contribution.

  The crucial point is that while the positive definite terms with
  couplings $\alpha_{q},\alpha^Q_{q}$ are proportional to
  $(y_{q}^{i})^2$, that is to the square of the entries in $\hat Y_q$,
  the negative terms with coupling $\nu_{q}$ are linearly proportional
  to $y_{q}^i$.  Then
  one vanishing diagonal entry in $\hat{Y}_{q}$ gets lifted to a
  nonzero value proportional to $\nu_{q}$. This entry gets aligned
  with the nonvanishing entries in
  $Z_{q,Q}$.  
    Written explicitly, we have
  $\hat{Y}_{q} \sim {\rm diag}\left(0,\nu_{q},v_{q}\right)$ together
  with $Z_{q}^{T}\simeq v_{Z_q} \left(0,1,0\right)$ and
  $Z_{Q}^{T}\simeq v_{Z_Q}\left(0,1,0\right)$.\footnote{Because of the attractive 
  nature of the $\nu_q$  term, $Z_Q$ and $Z_q$ get aligned with the second largest diagonal 
component of $\hat Y_q$,  which implies $c_Q, c_q \to  0$  and $s_Q, s_q \to  1$.} However, there is one
  more attractive operator, that is the determinant 
in the first term in~\eqn{eq:Vfph2},
  which favors a nonvanishing third entry in $\hat Y_q$.  A nonzero
  value proportional to the product $\mu_{q}\cdot \nu_{q}$ is thus
  induced, yielding $\hat{Y}_{q}\propto{\rm
    diag}\left(\mu_{q}\nu_{q}/v_q,\nu_{q},v_{q}\right)$.  In the end, if
  $\nu_{q}$ and $\mu_{q}$ are adequately small (that is
  $\nu_{q},\,\mu_{q} \ll v_q$) a hierarchical solution is obtained.
  With $\alpha^{Q}_{q}=\alpha_{q}=\lambda_{A_{q}}>0$ and the
  simplified choice $v_{Z_{Q}}=v_{Z_{q}}=v_{q}$ we obtain, at the
  minimum, the configuration $\hat{Y}_{q}\simeq v_{q}{\rm
    diag}\left(\epsilon_{q}',\epsilon_{q},1\right)$
  with~\cite{Espinosa:2012uu}\footnote{The analytical derivation of
    the result~\eqn{eq:hierarchy} is more easily carried out by
    assuming that the `lengths' of the vevs remains fixed at the
    values determined by $V_{\cal I}$ alone (respectively $v_{q}$,
    $v_{Z_q}$ and $v_{Z_Q}$).  In fact, $V_{\cal AR}$ and $V_{\cal A}$
    can induce shifts of order $\nu_{q}\epsilon_{q}/\lambda_q$ in the
    lengths.  However, this gives only negligible corrections to the
    hierarchical solution.}
\begin{eqnarray}
\epsilon_{q} & = & \frac{\lambda_{A_{q}}\nu_{q}/v_{q}}
{3\lambda_{A_{q}}^{2}-\mu_{q}^{2}/v_{q}^{2}},\qquad
\epsilon_{q}'=\frac{\mu_{q}/v_{q}}{\lambda_{A_{q}}}
\epsilon_{q},\qquad V_{q}^{{\rm min}}=-\nu_{q}\epsilon_{q}v_{q}^{3}\,.
\label{eq:hierarchy}
\end{eqnarray}
\Eqn{eq:hierarchy} seems to suggest that to obtain
e.g. $\epsilon_{q}'\sim10^{-4}$ and $\epsilon_{q}\sim10^{-2}$, in
rough agreement with the hierarchy in the up-quark Yukawa sector, a
mild parametric hierarchy like $\mu_{u}/v_{u}\sim \nu_{u}/v_{u}\sim
10^{-2}$ is needed (the milder hierarchy in the down-quark sector can be 
obtained  with  $ \mu_d/ v_d \sim \nu_d/ v_d \sim 10^{-1}$).  In fact, as we argue in the next section, after
coupling the up and down Yukawa sectors, a dynamical (as opposite to
parametric) suppression of $\epsilon'_q$ and $\epsilon_q$ arises.  The
only requirement to seed this suppression is $\mu_q, \nu_q \lsim v_q$,
which is, however, a natural requirement.  In fact, in the limit
$\mu_{q},\,\nu_{q}\to 0$ (that is $V_{\cal A}\to 0$), only Hermitian
monomials survive, and the potential acquires 
three  $U(1)$  symmetries  corresponding trivially to rephasing of the field multiplets $Z_Q$, $Z_q$ and $Y_q$. 
 This ensures that higher order corrections to $\mu_{q}$
and $\nu_{q}$ will be proportional to these same parameters, while in
contrast $v_q$ can receive larger corrections proportional for example
to the square of the cutoff scale $\Lambda$ in \eqn{eq:Ly}.  In
conclusion, the emergence of the large hierarchies observed in the SM
Yukawa sector is triggered by $\mu_q, \nu_q \lsim v_q$, which in turn
is what should be expected from simple naturalness considerations.




\section{Quark  mixings}
\label{sec:mixings}

In this section we extend the previous construction by including both
the $u$ and $d$ quark sectors. The flavor symmetry corresponds to the
full flavor group ${\cal G}_{\cal F}$.  As can be guessed from the
previous section, in order to obtain hierarchical solutions for both
the Yukawa fields $Y_{u}\sim\left(\mathbf{3},\mathbf{\bar
    3},\mathbf{1}\right)$ and
$Y_{d}\sim\left(\mathbf{3},\mathbf{1},\mathbf{\bar 3}\right)$, we need
to introduce three scalar multiplets transforming in fundamental
representations of the three group factors:
$Z_{Q_1}\sim\left(\mathbf{3},\mathbf{1},\mathbf{1}\right)$,
$Z_{u}\sim\left(\mathbf{1},\mathbf{3},\mathbf{1}\right)$ and
$Z_{d}\sim\left(\mathbf{1},\mathbf{1},\mathbf{3}\right)$. However, as
we will argue in the following, to get three nonvanishing quark
mixings we will need in fact to introduce one additional multiplet
$Z_{Q_2}\sim\left(\mathbf{3},\mathbf{1},\mathbf{1}\right)$.

Let us  write the matrices of  background Yukawa fields as 
\begin{eqnarray}
Y_{u} & = & \left(\vec{y}_{u1},\vec{y}_{u2},\vec{y}_{u3}\right),\nonumber \\
Y_{d} & = & \left(\vec{y}_{d1},\vec{y}_{d2},\vec{y}_{d3}\right),
\end{eqnarray}
where $\vec{y}_{ui}$ and $\vec{y}_{di}$ are column vectors with three
components, which we arrange according to
$|\vec{y}_{u1}|<|\vec{y}_{u2}|<|\vec{y}_{u3}|$ and
$|\vec{y}_{d1}|<|\vec{y}_{d2}|<|\vec{y}_{d3}|$.  Quark mixing can be
described as a misalignment in flavor space between the background
Yukawa matrices $Y_{u}$ and $Y_{u}$.  This is better understood if we
choose a special basis as follows: with $SU(3)_Q\times SU(3)_q$
rotations we can bring one $Y_q$ in diagonal form (with one complex
phase).  We choose to rotate $Y_u$ and we denote its diagonal matrix
as $\hat{Y}_{u}$, that is
$\vec{y}_{ui}\propto\left(\delta_{1i,}\delta_{2i},\delta_{3i}\right)^{T}$.
Transformations generated by the generators $\lambda_3^{Q+u}$,
$\lambda_8^{Q+u}$ and by the eight generators of $SU(3)_d$ leave
invariant $\hat{Y}_{u}$, and we can use this remaining freedom to
remove three moduli and seven phases from $Y_d$, which can thus be written as
$Y_{d}=K\hat{Y}_{d}$ with $K$ a special unitary matrix (with one
complex phase) and $\hat{Y}_{d}$ diagonal (with another complex
phase).  Clearly the matrix $K$
describes the misalignment between $Y_u$ and $Y_d$, and corresponds to
the CKM matrix.

\begin{figure}
 \includegraphics[scale=0.4]{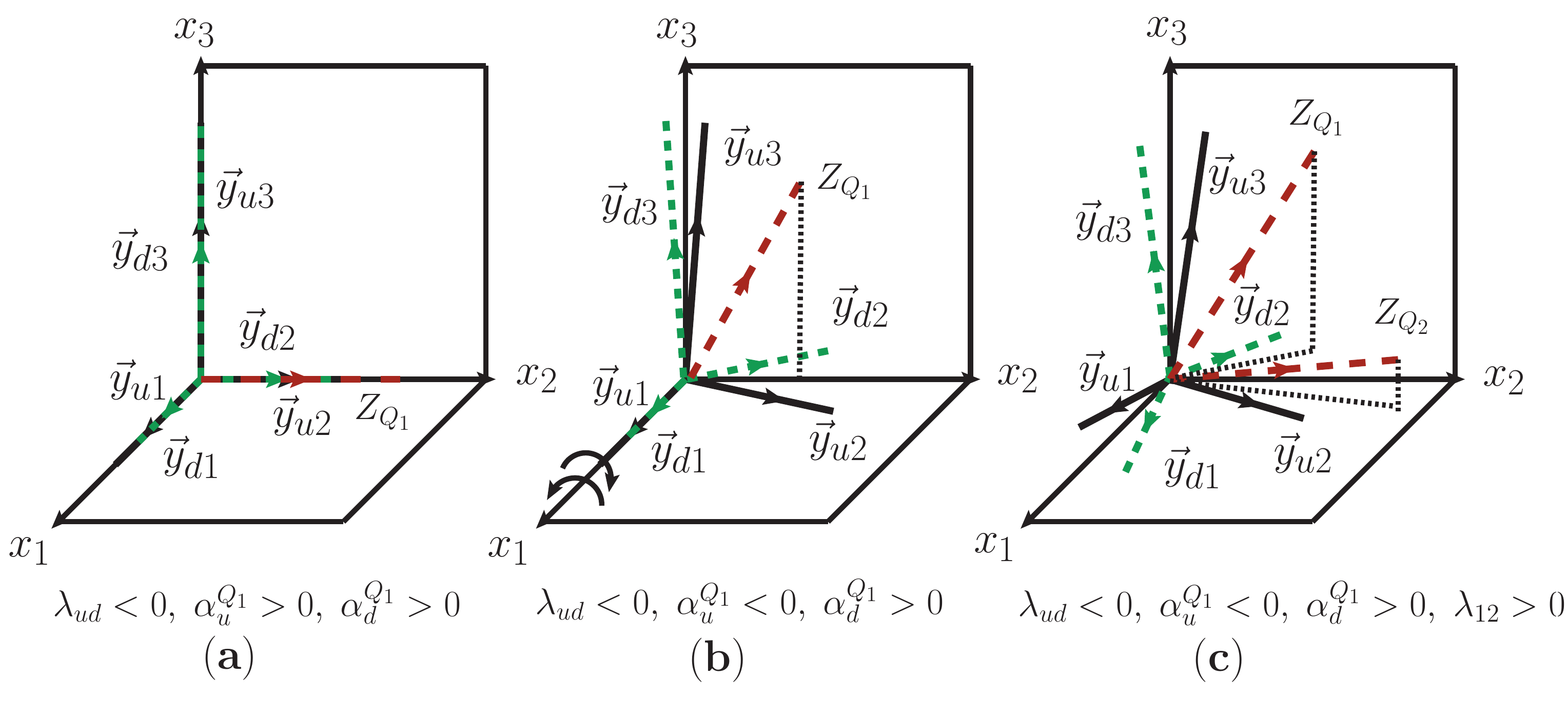}
 \caption{\label{fig:mixing}In (a), we illustrate the configuration with no
 mixing. In (b), we illustrate the configuration with one nonvanishing
 mixing angle $\theta_{23}$. In (c), we illustrate the configuration
 for nonvanishing $\theta_{12}$, $\theta_{23}$ and $\theta_{13}$.}
 \end{figure}

Since we will defer to the
next section the study of CP violation, for the time being we take all
the parameters of the scalar potential as well as all the background
field components to be real.  A simple graphical illustration can help to
understand how mixings can be induced.  Let us first draw the
vectors $\vec{y}_{ui}$ and $\vec{y}_{di}$ as in Fig. \ref{fig:mixing},
where the three axes $x_1$, $x_2$ and $x_3$ define a three-dimensional
flavor space, on which $Y_u$ and $Y_d$ get projected respectively
with components $\vec{y}_{ui}$ and $\vec{y}_{di}$.  Let us first
consider the $Y_u$-$Y_d$ interaction corresponding to the term
\begin{eqnarray}
&& \lambda_{ud}{\rm Tr} \left| \hat{Y}_{u}^{\dagger}K\hat{Y}_{d} \right|^2 .
\label{eq:lud}
\end{eqnarray}
This term, by itself, will align or maximally misalign ${Y}_{u}$ and
${Y}_{d}$. If $\lambda_{ud}<0$ the interaction is attractive, and the
largest possible values for the products between the components of
$Y_u$ and $Y_d$ are favored, which means that $\vec{y}_{u3}$ aligns
with $\vec{y}_{d3}$, $\vec{y}_{u2}$ with $\vec{y}_{d2}$ and
$\vec{y}_{u1}$ with $\vec{y}_{d1}$.  This configuration clearly yields
$K=I_{3\times3}$ that is no mixing, as is shown in Fig. 1(a).  If
$\lambda_{ud}>0$ the interaction is repulsive, and $Y_u$ and $Y_d$
will get maximally misaligned, that is the vector $\vec{y}_{u3}$
aligns with $\vec{y}_{d1}$ and $\vec{y}_{u1}$ with $\vec{y}_{d3}$
suppressing the two largest entries ($\vec{y}_{u2}$ remains aligned
with $\vec{y}_{d2}$).  Also in this configuration all the mixings
vanish, in the sense that $K$ becomes antidiagonal with unit entries.
In other words, in the basis in which the entries in $\hat Y_u$ and
$\hat Y_d$ are ordered in opposite ways, again we have
$K=I_{3\times3}$. Clearly, this means that the heaviest quarks get
coupled to the lightest ones, i.e.  $V_{td}=V_{ub}=1$, which conflicts
with observations.  We thus learn that $\lambda_{ud}<0$ is the correct
choice to get $K=I_{3\times3}$ as a first approximation and, as was
first noted in Ref. \cite{Anselm:1996jm}, that with just two Yukawa
fields $Y_u$ and $Y_d$, it is not possible to generate any mixing:
attractive interactions yield exact alignment, and repulsive
interactions result in maximal misalignment.

With $\lambda_{ud} <0$, the term in \eqn{eq:lud} besides aligning
$\hat Y_{u}$ and $\hat Y_{d}$, has also the important effect of
enhancing the hierarchies among their diagonal entries.  We have first
detected this effect in our numerical study, but there is a simple way
to understand the way it works.  Consider for the matrices $\hat
Y_{q}$ ($q=u,d$) the following two configurations:
\beqa \frac{1}{v} \hat Y_1 &=&
{\rm diag}\left(\delta',\delta,1\right), \\
\frac{1}{v} \hat Y_2 &=& 
{\rm diag}(\epsilon',\epsilon,1+\frac{1}{2}(\delta'^2+\delta^2))\,.
\eeqa
With $\epsilon',\epsilon\ll \delta',\delta\ll 1$, $\hat Y_1$ is mildly
hierarchical, while $ \hat Y_2$ is strongly hierarchical.  Up to
${\cal O}(\epsilon'^2,\epsilon^2)$ and ${\cal O}(\delta'^4,\delta^4)$
$\hat Y_1$ and $\hat Y_2$ are of equal length $\frac{1}{v^2} {\rm
  Tr}(\hat Y^\dagger_1\hat Y_1)=\frac{1}{v^2} {\rm
  Tr}(\hat Y^\dagger_2\hat Y_2)
=1+\delta'^2+\delta^2$. Consider now the contributions to the
potential at the minimum from the $\lambda_{ud}$  and  the
determinant terms. Setting for simplicity $\mu_u=\mu_d=\mu$ and neglecting
higher powers of the small parameters we have:
\beqa 
\frac{1}{v^4}\delta V(Y_1) &\sim& - \lambda_{ud} - 
\frac{4\mu}{v} \delta'\delta 
+{\cal O}(\delta^4), 
\\ 
\frac{1}{v^4}\delta V(Y_2) &\sim& 
- \lambda_{ud} -2\lambda{_{ud}}(\delta^2+\delta'^2)+{\cal
  O}(\epsilon'\epsilon)\,.  \eeqa
Clearly if $\frac{2\mu}{v} \lsim \lambda_{ud}$ the strongly
hierarchical configuration $\frac{1}{v} \hat Y_2 \sim {\rm
  diag}\left(\epsilon',\epsilon,1\right)$ gives the lowest minimum.
By carrying out a more detailed analysis including also higher powers
of $\epsilon'$ and $\epsilon$, one obtains that $\epsilon' \ll
\epsilon$ also lowers the minimum. The conclusion of this simple
discussion is that when the up and down sectors are coupled together,
the naive hierarchical pattern $y_{1}\sim \frac{\nu}{v}\mu$,
$y_{2}\sim \mu $ and $y_3\sim v$ derived at the end of the previous
section gets amplified by the effect of the $\lambda_{ud}$ term.
Numerically, we find (see section~\ref{sec:numerics}) that
$\frac{\mu}{v} \sim 10^{-1} \lambda_{ud}$ is  sufficient to
generate the strong hierarchies of the quark mass matrices. It is
quite remarkable that strong hierarchies can arise dynamically
from a potential in which the fundamental parameters are
nonhierarchical, and are only required to satisfy some 
generic naturalness condition.

Coming back to the issue of mixing angles, let us now include just the
minimal number of fields needed to generate the hierarchies, which is
$Z_{Q_1},\, Z_u$ and $Z_d$. We thus have in total 
three `vectors' $Z_{Q_1},Y_u$ and $Y_d$ transforming under  
the L-handed factor $SU(3)_Q$.
Looking at the terms which couple the $Z$'s and the $Y$'s, the
following ${\cal G}_{\cal F}$-invariant (attractive/repulsive) terms
are relevant to generate mixings:
\begin{eqnarray}
 \label{eq:alpha_QuQd}
 \alpha^{Q_1}_{u}\left|Z_{Q_1}^{\dagger}\hat{Y}_{u}\right|^2, 
& \qquad &  
\alpha^{Q_1}_{d} \left|Z_{Q_1}^{\dagger}K\hat{Y}_{d}\right|^2, \\
\alpha_{u}\left|Z_{u}^{\dagger}\hat{Y}_{u}^{\dagger}\right|^2, 
& \qquad & \alpha_{d} \left|Z_{d}^{\dagger}\hat{Y}_{d}^{\dagger}\right|^2.
\label{eq:alpha_ud}
\end{eqnarray}
In the second term in~\eqn{eq:alpha_QuQd} the matrix $K$ is sandwiched
between $Z_{Q_1}$ and $\hat Y_d$, and this implies that $Z_{Q_1}$
plays a direct role in determining the mixing pattern.  The roles of
$Z_{u}$ and $Z_{d}$ are instead indirect since they either do not
couple directly to $K$ as in~\eqn{eq:alpha_ud}, or they couple to
field combinations in which $K$ is sandwiched between the other two
fields, as for example in $(Z_{Q_1}^{\dagger}K\hat{Y}_{d})Z_{d}$ or in
$Z_{u}^{\dagger}(\hat{Y}_{u}^{\dagger}K\hat{Y}_{d})Z_{d}$.  These last
two monomials are non-Hermitian and thus always attractive, so that
they favor aligned configurations, for example between
$(Z_{Q_1}^{\dagger}K\hat{Y}_{d})$ and $Z_{d}$.  In short, the relevant
parameters which determine the mixing are $\alpha^{Q_1}_{u}$ which tends 
to align or maximally misalign $Z_{Q_1}$ and $\hat Y_u$, and $\alpha^{Q_1}_{d}$
which affects the structure of the matrix $K$.  If both
$\alpha^{Q_1}_{u}$ and $\alpha^{Q_1}_{d}$ are positive, $Z_{Q_1}$ will
repulse from both $Y_{u}$ and $Y_{d}$, a situation which is optimally
realized e.g. by the configuration $Z_{Q_1}^{T} =
v_{Z_{Q_1}}\left(0,1,0\right)$. Clearly in this case there is no
mixing (see Fig. \ref{fig:mixing}(a)).%
\footnote{Naively we would expect $Z_{Q_1}^{T}= v_{Z_{Q_1}}
  \left(c,s,0\right)$ (with $c^2+s^2=1$). However, the attractive
  terms $Z_{Q_1}^{\dagger}Y_{u}Z_{u}$ and
  $Z_{Q_1}^{\dagger}Y_{d}Z_{d}$, favor alignments with the second
  largest diagonal entries of $\hat{Y}_{u}$ and $\hat{Y}_{d}$.}  Now
if we switch the sign of one of the $\alpha^{Q_1}_q$ couplings,
something interesting occurs. Choosing for example
$\alpha^{Q_1}_{u}<0$ and $\alpha^{Q_1}_{d}>0$, there is attraction
between $Z_{Q_1}$ and $Y_{u}$ from the first term
eq. (\ref{eq:alpha_QuQd}), while there is repulsion between $Z_{Q_1}$
and $Y_{d}$ from the second term.  As a result, we obtain
$Z_{Q_1}^{T}=v_{Z_{Q_1}}\left(0,c,s\right)$ ($c^2+s^2=1$) as is
depicted in Fig. \ref{fig:mixing}(b). Notice that $Z_{Q_1}$ will
always lie in the $x_2-x_3$ plane because the attraction (repulsion)
with the component of $Y_{u}$ ($Y_{d}$) along the $x_{2,3}$-axis
dominates over the effects of the components along $x_1$. Due to the
attraction/repulsion, $Y_{u}$ and $Y_{d}$ will then rotate about the
$x_1$-axis, a nonzero $\theta_{23}$ is induced (see
Fig. \ref{fig:mixing}(b)), but the other two mixing angles remain
vanishing.

From this analysis, one can guess that to get three nonvanishing
mixing angles, at least four interacting ``vectors'' transforming
under the L-handed $SU(3)_Q$ flavor symmetry are need.  To generate
the other two mixing angles $\theta_{12}$ and $\theta_{13}$ let us
then introduce another scalar field $Z_{Q_2}$ transforming in the same
way as $Z_{Q_1}$ under $SU(3)_Q$.  Its interactions with $Y_{u,d}$
can be obtained directly from \eqn{eq:alpha_QuQd} by replacing
$Z_{Q_1}\to Z_{Q_2}$ and $\alpha^{Q_1}_{u,d}\to\alpha^{Q_2}_{u,d}$.
If both $\alpha^{Q_2}_{u,d}$ are positive, then the corresponding
terms will favor also in this case
$Z_{Q_2}^{T}\propto\left(0,1,0\right)$.  However, we also have the
following interactions involving the invariant
$Z_{Q_1}^{\dagger}Z_{Q_2}$:
\begin{eqnarray}
 &  & \lambda_{12}\left|Z_{Q_1}^{\dagger}Z_{Q_2}\right|^{2}
+\left[\tilde{\eta}_{12}\left(Z_{Q_1}^{\dagger}Z_{Q_2}\right)^{2}
+ {\cal M}^2_{12}\, 
Z_{Q_1}^{\dagger}Z_{Q_2}
%
+ {\rm h.c.}\right].
\label{eq:ZZp_terms}
\end{eqnarray}
In the above expression $ {\cal M}^2_{12}$ is a field dependent
quantity that reads
\begin{equation}
 {\cal M}^2_{12}= \tilde{\mu}_{12}^{2}  
+ \tilde{\eta}_u T_u  + \tilde{\eta}_d T_d \,, 
\label{eq:tildeM}
\end{equation}
where $ \tilde{\mu}_{12}^{2}, \tilde{\eta}_u, \tilde{\eta}_d$ are
complex parameters. It should be clear that small variations of ${\cal
  M}^2_{12}$ are uninfluential to determine the flavor structure, so
that in the minimization problem it is a consistent simplification 
approximating ${\cal M}^2_{12}$ with its background value
\begin{equation}
 \tilde m_{12}^2 \equiv \langle {\cal M}^2_{12}\rangle 
\simeq \tilde{\mu}_{12}^{2}  
+ \tilde{\eta}_u v_u^2  + \tilde{\eta}_d v_d^2 \,,  
\end{equation}
with $\tilde m^2_{12}$ a complex quantity.  The two non-Hermitian
monomials within square brackets in \eqn{eq:ZZp_terms} produce as
usual attractive interactions. If in addition $\lambda_{12}$ in the
first term is negative, $Z_{Q_2}$ will align with $Z_{Q_1}$. In this
case they will both remain on the $x_2-x_3$ plane and we do not get
any new mixing.  However, if $\lambda_{12}>0$ this term becomes
repulsive and if, as we will assume, its effect is the dominant one 
(which can be ensured by taking 
$ \lambda_{12} > |\tilde m_{12}^2| / (v_u^2 + v_d^2)$), 
then $Z_{Q_1}$ and $Z_{Q_2}$ can get sufficiently misaligned only if
they both leave the $x_2-x_3$ plane, as is illustrated in
Fig. \ref{fig:mixing}(c). Their couplings to $Y_{u}$ and $Y_d$ will
then induce rotations of the Yukawa matrices around the $x_2$ and
$x_3$ axes, with the result that nonvanishing values for $\theta_{13}$
and $\theta_{12}$ are generated (see Fig. \ref{fig:mixing}(c)).

In conclusion, while to obtain hierarchical solutions for both
$\hat{Y}_{u}$ and $\hat{Y}_{d}$ the set of auxiliary scalar fields
$Z_{Q_1}\sim\left(\mathbf{3},\mathbf{1},\mathbf{1}\right)$, 
$Z_{u}\sim\left(\mathbf{1},\mathbf{3},\mathbf{1}\right)$ and
$Z_{d}\sim\left(\mathbf{1},\mathbf{1},\mathbf{3}\right)$ is sufficient, this field content can
only generate one nontrivial mixing angle $\theta_{23}$.  In order to generate
the other two mixings $\theta_{12}$ and $\theta_{13}$ at least one
additional multiplet $Z_{Q_2}\sim\left(\mathbf{3},\mathbf{1},\mathbf{1}\right)$ is needed.  As is
well known, in the SM three nonvanishing mixing angles are a necessary
condition to allow for CP violation~\cite{Kobayashi:1973fv,Jarlskog:1985ht}.  
In the next section we will argue that the same
set of scalar multiplets is also sufficient to ensure that the ground
state of the scalar potential violates CP and induces a nonvanishing
complex phase in the quark mixing matrix $K$.
\section{CP violation}
\label{sec:CP}
We have seen that a scalar sector containing the two Yukawa fields
$Y_{u}$ and $Y_{d}$ and the four auxiliary multiplets
$Z_{Q_{1,2}}\sim\left(\mathbf{3},\mathbf{1},\mathbf{1}\right)$, 
$Z_{u}\sim\left(\mathbf{1},\mathbf{3},\mathbf{1}\right)$ and
$Z_{d}\sim\left(\mathbf{1},\mathbf{1},\mathbf{3}\right)$ 
can break the flavor group in such a
way that all the diagonal entries in the Yukawa matrices are
nonvanishing and naturally hierarchical, and moreover can misalign
$Y_u$ and $Y_d$ along all the three flavor directions inducing three
nonvanishing CKM mixing angles.  In deriving these results we have
taken for simplicity all quantities to be real.  In this section we
will complete our study by addressing the issue of CP violation and,
in order to do this, we will allow for complex parameters and complex
values of the background fields.  With the given field content the
most general renormalizable ${\cal G}_{\cal F}$-invariant potential
is:
\begin{equation}
V(Y_q,Z_q,Z_{Q_i}) =V_{\cal I}+V_{\cal AR}+V_{\cal A},
\label{eq:general_potential}
\end{equation}
where
\begin{eqnarray}
\label{eq:VI}
&& \hspace{-0.8cm} 
V_{\cal I}  =  
\sum_{q=u,d}\lambda_{q}\left(T_{q}-v_{q}^{2}\right)^{2}
 +\sum_{q=u,d}\lambda_{Z_{q}}\left(\left|Z_{q}\right|^{2}
-v_{Z_{q}}^{2}\right)^{2}
+\sum_{i=1,2}\lambda_{Q_i}\left(\left|Z_{Q_i}\right|^{2}-v_{Z_{Q_i}}^{2}\right)^{2} 
 \nonumber \\
&& \hspace{-1.cm} 
+ \ 
 \left\{\sum_{q=u,d} \left[g_q\left(T_{q}-v_{q}^{2}\right)+ g_{Z_q}\left(\left|Z_{q}\right|^{2}-v_{Z_{q}}^{2}\right)\right]
 +\sum_{i=1,2} g_{Q_i}\left(\left|Z_{Q_i}\right|^{2}-v_{Z_{Q_i}}^{2}\right)\right\}^2\!\!\!,
\\
&& \hspace{-0.8cm} 
\label{eq:VAR}
V_{\cal AR}  =  \sum_{q=u,d} \left(
\lambda_{A_{q}}A_{q} +\alpha_{q}\left|Z_{q}^{\dagger}Y_{q}^{\dagger}\right|^2 
+ \sum_{i=1,2}
\alpha^{Q_i}_{q}\left| Z_{Q_i}^{\dagger}Y_{q}\right|^2 \right)
 \nonumber \\
 && \hspace{-0.0cm} 
+ \  \lambda_{ud}{\rm Tr}\left|Y_{u}^{\dagger}Y_{d}\right|^2
  +   \lambda_{12} \left|Z_{Q_1}^{\dagger}Z_{Q_2}\right|^2, 
\\
&& \hspace{-0.8cm} 
\label{eq:VA}
V_{\cal A}  =  \sum_{q=u,d}\left[
 \widetilde{\mu}_{q}{\cal D}_{q}+
 \sum_{i=1,2}\widetilde{\nu}_{iq}
 Z_{Q_i}^{\dagger}Y_{q} Z_{q}\right] +
\widetilde{\gamma}_{ud}Z_{u}^{\dagger}Y_{u}^{\dagger}Y_{d}Z_{d} \nonumber \\
&& \hspace{-0.1cm} 
 + \  \widetilde{\eta}_{12}\left(Z_{Q_1}^{\dagger}Z_{Q_2}\right)^{2}
 +\widetilde{m}_{12}^{2}Z_{Q_1}^{\dagger}Z_{Q_2}
+{\rm h.c.}.
\end{eqnarray}
$V_{\cal A}$ contains nine complex couplings.\footnote{Dropping the
  simplification ${\cal M}^2_{12} \to \tilde{m}_{12}^{2} \equiv
  \langle {\cal M}^2_{12}\rangle$ we would have in fact two additional
  complex parameters $\tilde\eta_u$ and $\tilde\eta_d$, see
  \eqn{eq:tildeM}. However, the only role of the phases of these
  couplings is that of contributing to the overall phase of
  $\tilde{m}_{12}^{2}$ at the minimum. Adopting the simplified 
form of $V_{\cal A}$  \eqn{eq:VA} is thus justified.}
As shown in the Appendix, field redefinitions allow one to remove five
phases, leaving just four physical phases that can be chosen as the
two phase differences $\Delta_{q}$ $(q=u,d)$ between $\tilde\nu_{1q}$
and $\tilde\nu_{2q}$, the phase $\varphi_\gamma$ of the $(Y_u
Z_u)$-$(Y_d Z_d)$ coupling term, and the phase difference $\Delta_Q$
between the two complex parameters $\tilde\eta_{12}$ and $\tilde
m^2_{12}$ of the $Z_{Q_1}$-$Z_{Q_2}$ sector.  All the remaining
parameters in $V_{\cal A}$ can be taken, without loss of generality,
real and positive. In particular, the determinant terms can be written
as $\sum_{q} \mu_{q}{\cal D}_{q}+{\rm h.c.}$.

We address now the issue of weak CP violation. As is discussed in 
section  \ref{sec:mixings}, without loss of generality we can 
parametrize the vevs of the matrices of Yukawa fields as:
\beqa
Y_u &=& \hat Y_u, \\
Y_d &=& K\, \hat Y_d,
\eeqa
with $K$ a special unitary matrix depending on one phase $\delta_K$, and $\hat
Y_{u,d}$ both diagonal depending respectively on the two phases $\delta_{u,d}$.  
These latter two phases are fixed by the minimization
conditions for the determinants 
$\delta_{u,d}\to \pi$ so that at the minimum both $\hat Y_{u,d}$ are real. 
As regards the mixing matrix $K$, given that $\det K=+1$, 
the value of its phase $\delta_K$ is left undetermined by the 
minimization of the determinants, but will be  fixed 
after minimizing  the remaining terms in $V_{\cal A}$.
These terms can be rewritten as  (see \eqn{eq:phases} in the Appendix):
\beqa
\nonumber
V_{\cal A} &\supset& 
2 \sum_{i=1,2} \nu_{i u} \sum_{j} z_{Q_i}^j   y_u^{j} z_u^j 
\cos\left(\mp \Delta_u
-\phi_{Q_i}^j +\phi_{u}^j \right)\\
\nonumber
&+& 2 \sum_{i=1,2} \nu_{id} \sum_{jk} z_{Q_i}^j  K_{jk}\, y_d^{k} z_d^k 
\cos\left(\mp \Delta_d -\phi_{Q_i}^j +\xi^{jk}_{\delta_K}\right)\\
\nonumber
&+& 2\gamma_{ud} \sum_{jk} z_u^j y_u^j K_{jk} y_d^k z_d^k
\cos\left(\varphi_\gamma - \phi_{u}^j +\xi_{\delta_K}^{jk}\right) \\
\nonumber
&+& 2\eta_{12} \sum_{jk} z^j_{Q_1} z^j_{Q_2} z^k_{Q_1} z^k_{Q_2}
\cos\left(\Delta_Q - \phi_{Q_1}^j+\phi_{Q_2}^j- \phi_{Q_1}^k
+\phi_{Q_2}^k\right) \\
&+& 2 m^2_{12} \sum_j z^j_{Q_1} z^j_{Q_2}\cos\left(-\frac{1}{2}\Delta_Q
  - \phi_{Q_1}^j +\phi_{Q_2}^j\right), 
\label{eq:cosines}
\eeqa
where we denote the field components as $(\hat Y_q)^{jj} = y_q^j$,
$(Z_{Q_i})^{j}=z_{Q_i}^j e^{i\phi_{Q_i}^j}$,
$(Z_{q})^{j}=z_{q}^j e^{i\phi_{q}^j}$ and
in the first two lines the minus sign in front of $\Delta_{u,d}$
holds for $i=1$ and the plus sign for $i=2$. In \eqn{eq:cosines} the
quantities  $\phi^{jk}(\delta_K) \equiv   {\rm arg}(K_{jk})$ can be regarded  
as  functions of the phase $\delta_K$ of the mixing matrix.  
They always appear in the combinations $\xi^{jk}_{\delta_K}=\phi^{jk}(\delta_K)+
\phi_{d}^k$.  A few remarks are in order: \\ [.2cm]
\noindent
1.\ Although the functions $\phi^{jk}= \phi^{jk}(\delta_K)$ of
$\delta_K$ always come together with the phase of a $Z_d$ component,
these phase combinations satisfy $\xi^{jk}_{\delta_K}
+\xi^{lm}_{\delta_K}-\xi^{jm}_{\delta_K}- \xi^{lk}_{\delta_K} =
\phi^{jk}+ \phi^{lm} - \phi^{jm}- \phi^{lk} $, and thus the Jarlskog
invariant \cite{Jarlskog:1985ht} $J={\rm Im}(K_{jk} K_{lm} K^*_{jm}
K^*_{lk}) $ (no sum over repeated indices) is proportional to 
$ \sin\left(\phi^{jk}+ \phi^{lm} - \phi^{jm} -
  \phi^{lk}\right) $
and 
can be evaluated (numerically) without any
ambiguity. \\ [.2cm]
\noindent
2.\ There is a special choice of the phases that besides rendering
$\tilde\nu_{iq}, \tilde\gamma_{ud}, \tilde\eta_{12}$ and $ \tilde
m^2_{12}$ all real, also implies that at the minimum $J= 0$ and there
is no CP violation.  At fixed values of the moduli of the parameters,
this choice of phases corresponds to the lowest possible minimum.
This can be shown in a simple analytical way.  By means of the
redefinitions
\beq
Z_{q} \to e^{i\Delta_q} Z_q, \qquad
Z_{Q_2} \to e^{\frac{i}{2} \Delta_Q}Z_{Q_2}\, 
\eeq
$V_{\cal A}$ can be rewritten in the basis in which $m^2_{12}$ and
$\nu_{1q}$ are real (cf. \eqn{eq:phases}), while  the four complex
parameters become
\begin{eqnarray}
\label{eq:ph1}
\tilde \nu_{2q} &=&  \nu_{2q}\, e^{i(2 \Delta_q-\frac{1}{2}\Delta_Q)}, \\
\label{eq:ph2}
\tilde \gamma_{ud} &=& \gamma_{ud}\, e^{i(\varphi_{\gamma}- \Delta_u+\Delta_d)}, \\
\label{eq:ph3}
 \tilde \eta_{12} &=& \eta_{12}\, e^{2i\Delta_Q}. 
\end{eqnarray}
By choosing  the four phase combinations 
above 
all equal to $\pi$,  
it is easy to check  that  the minimum 
of $V_{\cal A}$ is obtained for 
\begin{eqnarray}
&& \phi^j_{Q_2}-\phi^j_{Q_1}  = \pi, \qquad  
\xi_{\delta_K}^{jk}= \phi_{u}^j=\phi^j_{Q_2}\,. 
\end{eqnarray}
Since in this case all the $\cos \to -1$, this corresponds to the best
possible minimum. Finally, given that $\phi^{jk}(\delta_K)$ can be
written as $ \phi^{jk}= \phi_{u}^j - \phi_{d}^k$, $J=0$ follows
straightforwardly. \\  [.2cm]
\noindent
3.\ Instead, we have not been able to prove analytically that for
generic values of $\Delta_q,\Delta_Q$ and $\varphi_\gamma$, $J\neq 0$ is
generally obtained. (Expressing $J$ as a function of the fundamental
phases would obviously be an even more awkward task.)  However, some
arguments can be put forth to suggest that this is indeed what should
be expected. Imagine for example to fix the value of $\delta_K$ to $0$
or $\pi$. There are several terms in $V_{\cal A}$ and, for generic
values of the phases, not enough field variables to drive all the
$\cos \to -1$. This is obvious for example for the last two lines in
\eqn{eq:cosines}, as well as in the first and in the second line once
the difference $\phi^j_{Q_2}-\phi^j_{Q_1}$ gets fixed in terms of
$\Delta_Q$. Minimization at fixed $\delta_K$ would then yield some
value $V_{\cal A}^{\rm min}$ higher than the best minimum of the
example above. Leaving now $\delta_K$ free, it is reasonable to expect
that some cosine term could be made smaller when the value of
$\delta_K$ departs from $0,\pi$, and a lower minimum could then be
reached.  In any case, the results of our numerical analysis confirm
that such an expectation is correct, and that this is precisely what
happens.


\section{A numerical example}
\label{sec:numerics}

In our study, the final verdict if spontaneous breaking of the flavor
symmetry is able to account for the entire set of observables in the
quark sector, has been settled only by means of numerical minimization
of the full ${\cal G}_{\cal F}$-invariant scalar potential. In
particular, we have not attempted to carry out multidimensional global
fits to the SM observables, which would have required a prohibitive
amount of CPU time, but we have just assumed a simple set of values
for most of the fundamental parameters and then, by varying the
remaining (crucial) ones, we have attempted to approximate the
experimental values of the observables. Of course, in carrying out
this procedure we have been guided by a good understanding of the role
of each term in the potential, which we have gained by inspecting
several partial analytical results.\footnote{Numerical minimizations
  have been carried out with the built-in minimization routines of the
  {\tt Mathematica} package.  To seek for the global minimum, we have
  used a {\it random search method}: we start with a random generation
  of initial search points in field space and proceed with the
  minimization routine. The minima of the potential resulting from
  different initial search points are compared, and the lowest one is
  selected. The set of initial search points is then augmented until
  there is no change in the final result.}  An example of the type of
results that can be obtained is given below.  We work in the basis in
which $\nu_{1q}$ ($q=u,d$) and $m_{12}^2$ are real and positive, and
the complex parameters are $\tilde\nu_{2q},\tilde\gamma_{ud}$ and
$\tilde\eta_{12}$.

In $V_{\cal I}$ we fix all the vevs to be equal
$v_q=v_{Z_q}=v_{Z_{Q_i}}=v$ and all the couplings to be 
equal to 1: $\lambda_q=\lambda_{Z_q}=\lambda_{Q_i}
=g_q=g_{Z_q}=g_{Q_i}=1$.   In $V_{\cal AR}$ we also fix
$\lambda_{A_q}=\alpha_q=\alpha^{Q_2}_q=\alpha^{Q_1}_d=1$ while 
\beq
\alpha^{Q_1}_u=-1,\qquad {\rm and}\qquad  \lambda_{ud}=-1.3\,. 
\label{eq:input1}
\eeq
In  $V_{\cal A}$  the dimensional parameters, the dimensionless 
couplings and the phases  are fixed as  ($q=u,d$) 
\begin{eqnarray}
 \mu_{q}=\nu_{1q}=\nu_{2q}= & v/10,\qquad  m_{12}^{2}= &0.15\, v^2, \qquad\nonumber \\
 \gamma_{ud}= & 0.81,\qquad \eta_{12}= &0.1, \qquad \lambda_{12}=1.27, \nonumber \\
 \phi_{\gamma_{ud}}= &0.98\pi, \qquad \phi_{\eta_{12}}=& 0.92\pi, \qquad  \phi_{\nu_{2q}}=0.95\pi.
\label{eq:input2}
\end{eqnarray}
With these inputs, the resulting parameters of the SM quark sector 
are
\begin{eqnarray}
|\hat{Y}_{u}| & = & v\>{\rm diag}\left(0.0003,0.009,1.4\right),
\nonumber \\
|\hat{Y}_{d}| & = & v\>{\rm diag}\left(0.0007,0.02,1.2\right),
\nonumber \\
K & = & \left(\begin{array}{ccc}
0.974 & 0.223 & 0.027\\
0.224 & 0.974 & 0.042\\
0.017 & 0.046 & 0.999
\end{array}\right),\nonumber \\
J & = & 2.9\times10^{-5}.
\label{eq:output1}
\end{eqnarray}
Let us note that having the largest entries in $\hat Y_{u,d}$ of
similar size, which follows from $v_d=v_u$, does not
constitute any problem. The value of the $b$-quark mass can be easily
suppressed by means of a $U(1)$ spurion vev, along the lines described
for example in \cite{Nardi:2011st}, or by extending the Higgs sector
to a two doublets model with $\langle H_d\rangle \ll \langle
H_u\rangle$.  Also, the fact that the phases in the last line of
\eqn{eq:input2} are all close to $\pi$, which implies somewhat small
imaginary parts for the complex parameters, is a simple consequence of
our attempt to reproduce the observed value of $J$ with an angle
$\theta_{13}$ a bit too large.  We include below for completeness also
the resulting values of the moduli of the auxiliary fields:
\begin{eqnarray}
|Z_{Q_1}^{T}| & = & v\left(0.05,0.10,0.98\right),\nonumber \\
|Z_{Q_2}^{T}| & = & v\left(0.04,0.79,0.05\right),\nonumber \\
|Z_{u}^{T}| & = & v\left(0.0001,0.70,0.37\right),\nonumber \\
|Z_{d}^{T}| & = & v\left(0.0007,0.66,0.43\right).
\end{eqnarray}

\section{Conclusions}
\label{sec:conclusions} 

In this paper we have shown that eight observables of the SM quark
sector (four mass ratios, three mixing angles and $\delta_K$) can be
reproduced by starting from the simple idea that the complete breaking
of the quark flavor symmetry results as the dynamical effect of
driving a suitable scalar potential towards its minimum. We have
identified the minimum set of multiplets in simple (fundamental and
bifundamental) representations of the group needed to break ${\cal
  G}_{\cal F}\to 0$, and we have shown that this same set of fields is
also sufficient to generate one weak CP violating phase. Besides the
quantitative results, through this study we have gained important
qualitative understandings of various mechanisms that might underlie
some of the most puzzling features of the SM quark sector. We list
them in
what we think is their order of importance. \\  [-0.2cm]

1. $K = V_{CKM} \approx I_{3\times 3}$.\ The interaction between the two
Yukawa fields $Y_u$ and $Y_d$ tends to generate an exact alignment of
their vevs in flavor space, resulting in $V_{CKM} = I_{3\times 3}$ 
\cite{Anselm:1996jm}.
If the interaction is repulsive ($\lambda_{ud} > 0$) the alignment
occurs when the eigenvalues of the two matrices are ordered by size in
an opposite way.  When the interaction is attractive ($\lambda_{ud} <
0$) the alignment occurs when the ordering is the same. This second
possibility is the one observed in nature. To generate three
nonvanishing mixing angles, that is to (slightly) misalign $Y_u$ and
$Y_d$ in all flavor directions, at least two other multiplets
transforming under the L-handed factor $SU(3)_Q$ are needed.  Their
presence will induce perturbation in the exact alignment, but if the
$Y_u$-$Y_d$ interaction is sufficiently strong, $V_{CKM} \approx
I_{3\times 3}$ will be maintained.\\  [-0.2cm]

2. {\it Yukawa hierarchies.}\ Hierarchies between the different
entries in $Y_u$ and $Y_d$ are seeded by taking for a subset of the
dimensional parameters values somewhat smaller than the overall scale
of the vevs: $\mu_q, \nu_{1q},\nu_{2q}\sim v/10$. This can be
justified by the fact that when these parameters are set to zero, the
scalar potential gains some additional $U(1)$ invariances.  The initial
(mild) suppression of some entries in $Y_{u,d}$ can get enhanced down
to the observed values of the quark mass ratios by dynamical effects.
Hierarchical Yukawa couplings can then be generated without
strong hierarchies in the fundamental parameters.  \\  [-0.2cm]

3. {\it Weak CP violation.}\ Once the flavor symmetry is completely
broken, generating the CKM CP violating phase does not require
complicating further the model.  The set of scalar multiplets needed
to obtain ${\cal G}_{\cal F}\to 0 $ ensures that several complex
phases cannot be removed regardless of field redefinitions, and this
ensures that the scalar potential contains CP violating terms.  For
generic values of these phases, a CP violating
ground state for $Y_{u,d}$ is obtained.  \\  [-0.2cm]

Indeed, one could object that in our construction there are many more
fundamental parameters than there are observables. This of course
affects its predictivity, and in some respects also its elegance.  We
cannot object to such a criticism, but it is worth stressing that the
proliferation of parameters is a result of the usual quantum field theory prescription
for building renormalizable Lagrangians: we have identified the
minimum number of multiplets needed to break completely ${\cal
  G}_{\cal F}$, and next we have simply written down the complete set
of renormalizable operators allowed by the symmetry. After all, as it
has been argued e.g.  in~\cite{Duque:2008ah}, the apparent lack of
simple relations between the observables of the quark sector might well
be due to the fact that, as in our case, they are determined by a
very large number of fundamental parameters.

Direct evidences of  the scenario we have been studying might arise from the fact that if the 
flavor symmetry is global, then spontaneous symmetry breaking implies the presence of
Nambu-Goldstone bosons that could show up in yet unseen hadron decays or in rare flavor
violating processes. If the flavor symmetry is instead gauged, then to ensure the absence of 
gauge anomalies additional fermions must be introduced~\cite{Grinstein:2010ve}, 
and their detection could then represent a smoking gun for this type of models. All this remains, 
however, a bit speculative, especially because the theory provides no hint of the scale at which the
 flavor symmetry gets broken, and very large  scales would suppress most, if not all, types of signatures.

\section*{Note added}

In ref.~\cite{Fong:2013sba} we put forth the idea that the spontaneous
breaking of the quark-flavor symmetry could automatically solve the
strong CP problem. The mechanism underlying this idea was that, after
rotating $\theta_{QCD}$ and all other potentially dangerous phases
into the scalar potential, at the minimum the vevs of the Yukawa field
matrices satisfy ${\rm Arg} \left[\det \left(\langle {Y}_u\rangle
    \langle {Y}_d\rangle \right)\right] = 0$ $({\rm mod}\
2\pi)$. While this is true, when the Yukawa field vevs are reinserted
into the effective operators \eqn{eq:Ly}, unremovable phases (and in
particular $\theta_{QCD}$) reappear in the Yukawa matrices $ {\cal
  Y}_u$ and ${\cal Y}_d$ (defined after ref.~\eqn{eq:Ly}).  Therefore, the
claim made in~\cite{Fong:2013sba} is incorrect. The issue whether
spontaneous breaking of the quark-flavor symmetry can provide some
alternative mechanism to solve the strong CP problem is presently
under investigation.

\section*{Appendix: Physical phases of the scalar  potential}

The complex parameters of the scalar potential 
eq. (\ref{eq:general_potential}) all appear in $V_{\cal A}$
\eqn{eq:VA} which can be rewritten as: 
\begin{eqnarray}
V_{\cal A}  &=&  \sum_{q=u,d}\left[
 e^{i\phi_{\mu_q}}{\mu}_{q}{\cal D}_{q}+
 \sum_{i=1,2} e^{i\phi_{\nu_{iq}}} \nu_{iq}
 Z_{Q_i}^{\dagger}Y_{q} Z_{q}\right] + 
e^{i\phi_{\gamma_{ud}}}\gamma_{ud}Z_{u}^{\dagger}Y_{u}^{\dagger}Y_{d}Z_{d}  
\nonumber \\ &+&
e^{i\phi_{\eta_{12}}} {\eta}_{12}\left(Z_{Q_1}^{\dagger}Z_{Q_2}\right)^{2} 
+ e^{i\phi_{m^2_{12}}} m_{12}^{2}Z_{Q_1}^{\dagger}Z_{Q_2} 
+{\rm h.c.}\,,
\label{eq:appVA}
\end{eqnarray}
where in the last term we have absorbed the vacuum expectation value
of $\sum_{q} \tilde\eta_{q} T_q \approx \sum_{q}
\tilde \eta_{q} v^2_q $ ($q=u,d$).
Let us redefine the fields as follows
\begin{eqnarray}
Y_{q} & \to & e^{-\frac{i}{3}\phi_{\mu_{q}}}Y_{q}, 
\nonumber \\
Z_{q} & \to & e^{-\frac{i}{2}(
\phi_{\nu_{1q}}+\phi_{\nu_{2q}}-\frac{2}{3}\phi_{\mu_q})}Z_{q},
\nonumber \\
Z_{Qi} & \to & e^{\pm \frac{i}{8} \left(\phi_{\eta_{12}}+
2 \phi_{m^2_{12}}\right)}Z_{Qi}.
\end{eqnarray}
where  in the last line the plus sign
is for $Z_{Q_1}$ and the minus sign for $Z_{Q_2}$. After these redefinitions 
\eqn{eq:appVA} becomes: 
\begin{eqnarray}
V_{\cal A}  &=&  \sum_{q=u,d}\left[
 {\mu}_{q}{\cal D}_{q}+ 
\left(e^{-i \Delta_q}\,  \nu_{1q} Z_{Q_1}^{\dagger}
+ 
e^{i  \Delta_q}\,  \nu_{2q} Z_{Q_2}^{\dagger}\right) Y_{q} Z_{q}\right]\qquad\qquad\qquad\quad 
\nonumber \\
 && \hspace{-1.5cm} + \>
e^{i\varphi_\gamma}\,\gamma_{ud}Z_{u}^{\dagger}Y_{u}^{\dagger}Y_{d}Z_{d} + 
e^{i\Delta_Q} {\eta}_{12}\left(Z_{Q_1}^{\dagger}Z_{Q_2}\right)^{2} + 
e^{-\frac{i}{2}\Delta_Q} m_{12}^2Z_{Q_1}^{\dagger}Z_{Q_2} +{\rm h.c.},
\label{eq:phases}
\end{eqnarray}
where in terms of the initial phases in \eqn{eq:appVA}, 
we have $\Delta_q= \frac{1}{2}\left(\phi_{\nu_{2q}}-\phi_{\nu_{1q}}\right)
+\frac{1}{8}\left(\phi_{\eta_{12}}+ 2 \phi_{m^2_{12}}\right)$
(for $q=u,d$), $\Delta_Q=
\frac{1}{2}\phi_{\eta_{12}} - \phi_{m^2_{12}}$ and 
$\varphi_\gamma=\phi_{\gamma_{ud}}+\frac{1}{2}\left(
  \phi_{\nu_{1u}}-\phi_{\nu_{1d}}+
  \phi_{\nu_{2u}}-\phi_{\nu_{2d}}\right)$.


\end{document}